\documentclass[11pt]{article}

\usepackage[T1]{fontenc}
\usepackage[utf8]{inputenc}
\usepackage{textcomp}
\usepackage{lmodern}
\usepackage[margin=1in]{geometry}
\usepackage{amsmath,amssymb}
\usepackage{bm}
\usepackage{braket}
\usepackage{graphicx}
\usepackage{booktabs}
\usepackage{array}
\usepackage{booktabs}
\usepackage{caption}
\usepackage{float}      

\usepackage{enumitem}
\usepackage{authblk}
\usepackage{xcolor}
\usepackage[normalem]{ulem}
\usepackage[super,sort&compress,comma]{natbib}
\usepackage{microtype}
\usepackage[hidelinks]{hyperref}
\usepackage{listings}

\graphicspath{{figures/}}

\title{Sample-Based Quantum Diagonalization for Variable Molecular Geometries}

\author[1]{Matthieu Fortin-Deschênes\thanks{Email: \href{mailto:matthieu.fortin-deschenes@pinq2.com}{matthieu.fortin-deschenes@pinq2.com}}}
\author[2]{Ibrahim Shehzad}
\affil[1]{PINQ\textsuperscript{2}, Sherbrooke, QC J1K 1B7, Canada}
\affil[2]{IBM Quantum, Thomas J.\ Watson Research Center, Yorktown Heights, NY 10598, USA}
\date{}

\begin{document}

\maketitle

\begin{abstract}
Quantum and hybrid quantum-classical algorithms have the potential to improve the accuracy and scale of quantum chemistry and impact a broad range of applications such as drug discovery and catalysis. In particular, subspace methods such as sample-based quantum diagonalization (SQD) can already tackle classically challenging systems. However, practical algorithms that go beyond the computation of electronic ground and excited states are still scarce. Here, we introduce the variable-geometry sample-based quantum diagonalization (VGSQD) algorithm, which extends SQD to variable molecular geometries by reusing quantum samples from a reference geometry at displaced ones, thus significantly reducing quantum computational cost. Using circuits of up to 124 qubits on \texttt{ibm-quebec}, an IBM Heron \texttt{r2} quantum processor, we determine the equilibrium geometry and energy of the weakly bound, strongly correlated beryllium dimer, as well as the water molecule. Finally, we extend VGSQD to chemical reactions and compute the minimum energy path and energy barrier of a hydrogen exchange reaction. The results are comparable with state-of-the-art classical methods such as CCSD(T) and DMRG. This work broadens the scope of quantum algorithms for chemistry, from energies to geometry optimization and reaction pathways, both essential for practical applications.
\end{abstract}

\section*{Introduction}

Quantum chemistry is widely regarded as one of the most promising applications of quantum computing. Quantum algorithms such as quantum phase estimation\cite{kitaev1995,aspuru2005} and hybrid algorithms such as the variational quantum eigensolver\cite{peruzzo2014,mcclean2016,kandala2017} emerged first as important building blocks for future fault-tolerant and near-term devices, respectively. While noise hinders the application of the former to problems of a meaningful size, the latter is hampered by measurement overhead and optimization difficulties. On the other hand, subspace methods recently emerged as leading candidates for near-term and early fault tolerant devices, owing to their applicability to problems requiring large number of qubits, and their potential to improve together with the quantum hardware\cite{nakagawa2024,motta2024}. By preparing and probing a quantum state with overlap with the ground state on a Quantum Processing Unit (QPU), these methods allow the construction of a subspace onto which the Hamiltonian is projected and diagonalized at a reduced classical cost. Algorithms such as sample-based quantum diagonalization (SQD)\cite{robledo2025}, which builds on quantum-selected configuration interaction \cite{dmn4-snfx} have been particularly successful. In fact, together with algorithms that leverage them\cite{piccinelli2025,shajan2025,yu2025,piccinelli2026,merz2026}, they already allow for ground and excited state calculations of classically challenging molecules and have been applied to several use cases\cite{merz2025,kaliakin2025,barison2025,liepuoniute2025,smith2025,shivpuje2025,yamamoto2026}, such as solvated molecules and larger biological molecules.

While subspace methods have demonstrated their capability for ground and excited state computations, it is critical to develop algorithms that enable their application to practical quantum chemistry workflows. For instance, Kaliakin et al.\ demonstrated the use of implicit solvent methods to model molecule-solvent interactions\cite{kaliakin2025}. Another key component of chemistry workflows is the modeling of molecules with variable nuclear coordinates. This is necessary to carry out structural relaxation of molecules to find their equilibrium geometry, as well as to find the minimum energy path (MEP) and energy barrier of chemical reactions. In a naive SQD approach, each molecular geometry requires its own round of circuit execution and sampling, thus the QPU cost grows with the number of geometries visited and the workflow involves frequent alternation between quantum and classical resources. To reduce the number of required quantum circuit executions, we developed the variable geometry sample-based quantum diagonalization (VGSQD) algorithm. Analogous to correlated sampling in quantum Monte Carlo, which uses sampled configurations of a reference at displaced geometries\cite{filippi2000correlated}, VGSQD reuses quantum samples for multiple molecular geometries, thus significantly reducing the number of quantum circuit executions.

First, we describe how VGSQD extends SQD by reusing bitstrings sampled on a QPU across multiple closely related molecular geometries, and we introduce two criteria to assess when such reuse is justified. We then apply VGSQD to the structural relaxation of small molecules using a quantum-centric supercomputing workflow leveraging an IBM Heron \texttt{r2} quantum processor (\texttt{ibm-quebec})  and a high-performance computing cluster (\textit{Harfang}). The computations are done using the cc-pVDZ and cc-pVTZ basis sets, with circuits of up to 62 and 124 qubits respectively.  We first consider Be$_2$, whose unusual bonding makes its equilibrium bond length notoriously difficult to model. Geometry optimization with VGSQD yields relaxed bond lengths that are closer to the experimental value than those obtained with CCSD(T) for both basis sets. As for the ground state energy, the value for the relaxed Be$_2$ molecule is lower than DMRG and CCSD(T) for cc-pVDZ and on par with CCSD for cc-pVTZ. We then consider H$_2$O, which has more nuclear degrees of freedom to optimize. With the larger basis set (cc-pVTZ), the bond lengths and bond angles converge within 0.002 Å and 0.5° of the experimental values. Finally, we demonstrate the application of VGSQD to the study of chemical reactions by combining it with the climbing-image nudged elastic band (CI-NEB)\cite{jonsson1998,henkelman2000a,henkelman2000b} method to compute the minimum energy path and energy barrier of the $\mathrm{H_2} + \mathrm{H} \rightarrow \mathrm{H} + \mathrm{H_2}$ hydrogen exchange reaction. The obtained saddle point geometry and barrier height are within 0.2 kcal/mol and 0.001 Å of the FCI reference, thus demonstrating the potential of VGSQD to broaden the scope of quantum computing algorithms to chemical reactions on near-term quantum hardware.

\section*{Methods}

We work within the Born-Oppenheimer approximation and represent the electronic Hamiltonian in second quantization in the basis of Hartree-Fock molecular orbitals (MOs) $\phi_p$ obtained from either restricted Hartree-Fock (RHF) for closed-shell systems (Be$_2$ and H$_2$O) or restricted open-shell Hartree-Fock (ROHF) for open-shell systems ($\mathrm{H_2} + \mathrm{H} $), computed with PySCF\cite{sun2018}. The lowest $n_{\mathrm{frozen}}$ MOs are kept doubly occupied and frozen, and the calculation is carried out in the active space spanned by the remaining $N$ orbitals. In chemists' notation for the two-electron integrals, $(pq|rs) \equiv \iint \phi_p^*(\mathbf{r}_1)\,\phi_q(\mathbf{r}_1)\, r_{12}^{-1}\, \phi_r^*(\mathbf{r}_2)\,\phi_s(\mathbf{r}_2)\, d\mathbf{r}_1\, d\mathbf{r}_2$, the Hamiltonian reads:
\begin{equation}
\hat{H} = E_{\mathrm{core}} + \sum_{pq \in \mathrm{active}} \sum_{\sigma} h_{pq}^{\mathrm{eff}}\, \hat{a}_{p\sigma}^{\dagger} \hat{a}_{q\sigma} + \frac{1}{2} \sum_{pqrs \in \mathrm{active}} \sum_{\sigma\tau} (pq|rs)\, \hat{a}_{p\sigma}^{\dagger} \hat{a}_{r\tau}^{\dagger} \hat{a}_{s\tau} \hat{a}_{q\sigma},
\end{equation}
where $a_{xy}$ ($a_{xy}^{\dagger}$) represent the annihilation (creation) operators, $p,q,r,s$ label spatial MOs, $\sigma,\tau \in \{\alpha,\beta\}$ label spin, $h_{pq}^{\mathrm{eff}}$ are the effective active space one-electron (kinetic plus electron-nuclear attraction) integrals\footnote{Note that we do not freeze any molecular orbitals anywhere in this study.}, and $E_{\mathrm{core}}$ is the nuclear repulsion energy, including the frozen core energy.

We seek the ground state, $\ket{\Psi_0}$, of this Hamiltonian, as a linear combination of Slater determinants $\ket{\Phi_I}$ (with CI coefficients $c_I$), which form a basis for the antisymmetric N-fermion Hilbert space:
\begin{equation}
\ket{\Psi_0} = \sum_{I} c_I \ket{\Phi_I}
\end{equation}

\subsection*{Quantum circuits and diagonalization}

For ground state calculation of the reference molecule, we use the sample-based quantum diagonalization approach. We first prepare $\ket{\Psi_{\mathrm{QC}}}$, a state whose support in the computational basis overlaps with the support of the true ground state $\ket{\Psi_0}$ of the reference molecule. We then sample from it, i.e., we draw the set $\tilde{\chi}=\lbrace$$x$|$x$$\sim$$\tilde{P}$($x$)$\rbrace$, where $x \in \{0,1\}^{2N}$ are bitstrings distributed according to the noisy distribution $\tilde{P}(x)$, which approximates the ideal distribution $P(x) = |\braket{x|\Psi_{\mathrm{QC}}}|^2$ in the noiseless limit. Fermionic modes are mapped to qubits with the Jordan-Wigner transformation\cite{jordan1928}, using $2N$ qubits for $N$ active spatial orbitals. The first $N$ qubits encode the occupations of the spin-$\beta$ orbitals and the last $N$ qubits those of the spin-$\alpha$ orbitals, so that each measured bitstring corresponds directly to a Slater determinant of the active space.

To prepare the sampled state, we use the truncated local unitary cluster Jastrow (LUCJ) ansatz\cite{neuscamman2013,matsuzawa2020,motta2023}:
\begin{equation}
\ket{\Psi_{\mathrm{QC}}} = \hat{U} \ket{x_{\mathrm{RHF}}} = e^{\hat{K}_2} e^{\hat{K}_1} e^{i\hat{J}_1} e^{-\hat{K}_1} \ket{x_{\mathrm{RHF}}}
\end{equation}
where $\ket{x_{\mathrm{RHF}}}$ is the restricted Hartree-Fock determinant (ROHF for open-shell), $\hat{K}_\mu$ are one-body operators, whose exponentials are orbital rotations and $\hat{J}_\mu$ are density-density operators, as defined by:
\begin{equation}
\hat{K}_\mu = \sum_{pq,\sigma} K_{pq}^{\mu}\, \hat{a}_{p\sigma}^{\dagger} \hat{a}_{q\sigma}, \qquad
\hat{J}_\mu = \sum_{pq,\sigma\tau} J_{p\sigma,q\tau}^{\mu}\, \hat{n}_{p\sigma} \hat{n}_{q\tau}
\end{equation}

The ansätze are prepared using the ffsim library \cite{sung2026ffsim}  and their parameters are obtained from CCSD calculations, with no further optimization, following previous SQD studies\cite{robledo2025,piccinelli2025}. For closed-shell molecules the spin-balanced form ($J_{\alpha\alpha} = J_{\beta\beta}$, $J_{\alpha\beta} = J_{\beta\alpha}$) is used; for open-shell molecules, the spin-unbalanced form with independent $J_{\alpha\alpha}$, $J_{\beta\beta}$ and $J_{\alpha\beta}$ and ROHF-based CCSD amplitudes are used. The  $\alpha$ and $\beta$  spin orbitals are each mapped to individual qubit chains, with both chains coupled together through ancilla qubits. The 156 qubit IBM Heron \texttt{r2} chip with heavy-hex topology can accommodate two parallel chains with at most 26 qubits each. Longer chains must include a non-parallel section. For this reason, the spin orbitals of molecules with fewer than 26 MOs (52 spin orbitals) are mapped to $\alpha$ and $\beta$ qubit registers that form two parallel chains connected by ancillas at every fourth qubit. For molecules with more than 52 spin orbitals, the two spin registers are mapped onto chains with parallel sections (longest available on the chip) and non-parallel sections. The lowest energy MOs are assigned to those parallel sections and the remaining MOs are mapped to the two non-parallel segments. Among all embeddings of these patterns into the coupling graph of the quantum chip, the one minimizing the summed two-qubit gate and readout errors of all qubits of the circuit is selected. The locality constraint of the LUCJ ansatz restricts nonzero elements of the Jastrow matrices to pairs of orbitals whose qubits are adjacent, thus conserving all $J_{p\sigma,p+1\sigma}$ same spin interaction terms and opposite-spin $J_{p\sigma,p\tau}$ terms with rungs connecting the parallel sections of the chains (every fourth orbital of these sections).

\subsection*{Molecular orbital alignment}

Significant QPU computational cost is associated with running different circuits for different molecular geometries at each iteration of structural relaxation or MEP calculation. Furthermore, the interdependency between the QPU and classical workloads prevents their parallel execution and thus introduces latency as jobs are queued for QPU or classical HPC execution at each iteration. It is therefore highly desirable to reuse bitstrings sampled from the circuits of a given molecular geometry for multiple iterations with slightly different molecular geometries.

Here, we label the reference molecule as \emph{A} and its slightly distorted counterpart as \emph{B}. The molecular orbitals defining the Hartree-Fock ground states of the two geometries $X = \{A,B\}$ in their respective atomic orbital bases are:
\begin{equation}
\ket{\phi_q^{(A)}} = \sum_{\nu} C_{\nu q}^{(A)} \ket{\chi_\nu^{(A)}}, \quad \text{and} \quad
\ket{\phi_p^{(B)}} = \sum_{\mu} C_{\mu p}^{(B)} \ket{\chi_\mu^{(B)}}
\end{equation}
And the molecular orbital bases overlap matrix between the two geometries is defined by:
\begin{equation}
S_{\mathrm{MO}}^{(BA)} = C^{(B)\dagger} S_{\mathrm{AO}} C^{(A)}
\end{equation}
With atomic-bases overlap matrix elements given by:
\begin{equation}
S_{\mathrm{AO},\mu\nu}^{(BA)} = \braket{\chi_\mu^{(B)} | \chi_\nu^{(A)}}
\end{equation}

Since the Hartree-Fock reference is invariant under unitary rotations within the occupied and virtual spaces (and, for open-shell molecules, within the singly occupied space), individual molecular orbitals are not uniquely defined and may undergo arbitrary rotations as the molecular geometry changes. Consequently, orbital identities may vary significantly and even discontinuously, despite the underlying electronic state varying smoothly. Bitstrings sampled from the circuits constructed for the reference geometry can therefore not be reused directly for its distorted counterparts.

In order to ensure that the bitstrings from the distorted geometry correspond to those sampled for the reference geometry, we rotate\cite{king1967,olsen1995} the MOs of the frozen, doubly occupied, singly occupied and virtual blocks of molecule \emph{B}, i.e.\ $C_{\mathrm{occ}}^{B} \rightarrow C_{\mathrm{occ}}^{B} U_{\mathrm{occ}}$ and $C_{\mathrm{virt}}^{B} \rightarrow C_{\mathrm{virt}}^{B} U_{\mathrm{virt}}$, where the subspace rotations $U_{\mathrm{occ}}$ and $U_{\mathrm{virt}}$ are chosen as:
\begin{equation}
\min_{U} \left( \sum_{i}^{N_{\mathrm{subspace}}} \left\| \ket{\phi_i^{(A)}} - \ket{\tilde{\phi}_i^{(B)}} \right\|^2 \right)
\end{equation}
where $\ket{\tilde{\phi}_i^{(B)}}$ are the rotated MOs for each subspace. This problem is equivalent to maximizing the traces $\mathrm{Tr}\big(S_{\mathrm{MO}}^{(AB)} U\big)$ in the occupied and virtual subspaces and has the solution:
\begin{equation}
U = U_R U_L^{T}
\end{equation}
where the matrices $U_R$ and $U_L$ are obtained from the singular value decompositions in each subspace:
\begin{equation}
S_{\mathrm{MO}}^{(AB)} = U_L \Sigma U_R^{T}
\end{equation}

\subsection*{Nuclear gradients}

Nuclear gradients are computed semi-numerically by central finite differences of the ground state energy at fixed one- and two-electron reduced density matrices (RDMs) $\gamma$ and $\Gamma$, following Coe et al.\cite{coe2022,coe2023}. The gradient components are given by:
\begin{equation}
\frac{\partial E}{\partial R_{I\alpha}} \approx \frac{1}{2h} \left[ \sum_{pq} \gamma_{pq}\, \Delta h_{pq} + \frac{1}{2} \sum_{pqrs} \Gamma_{pqrs}\, \Delta (pq|rs) + \Delta E_{\mathrm{nuc}} \right]
\end{equation}
Here the $R_{I\alpha}$ are the $x$, $y$, $z$  coordinates of nucleus $I$. Hartree-Fock is solved at both displaced geometries of the finite differences evaluations, and the resulting MOs are aligned to the reference orbitals as described above. The one- and two-electron integrals are transformed to the aligned MO bases, and the gradient component is evaluated with the RDMs of the current state held fixed.

\subsection*{VGSQD algorithm overview}

\begin{figure}[htbp]
\centering
\includegraphics[width=\linewidth]{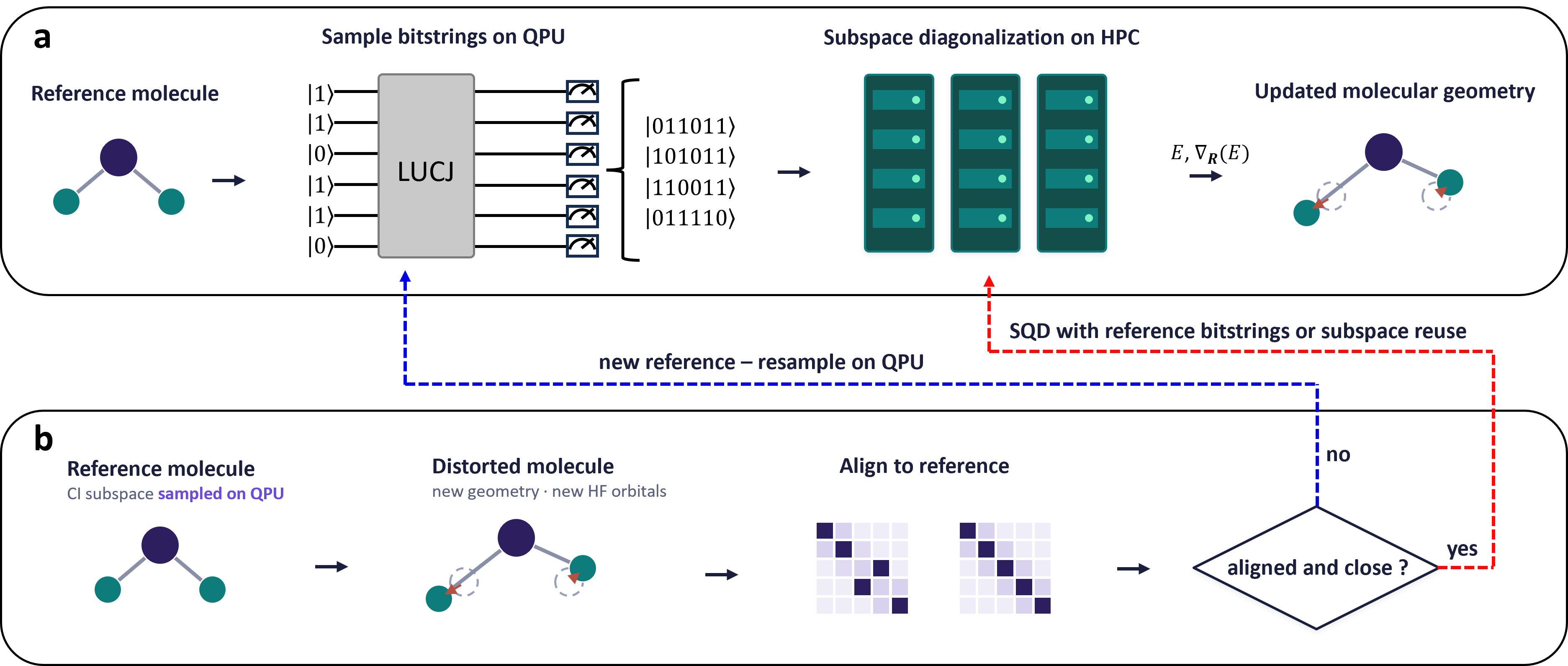}
\caption{Schematic of the VGSQD algorithm. (a) VGSQD workflow: LUCJ circuit sampling on the QPU for the reference molecule, subspace diagonalization on HPC and geometry update. (b) Alignment of the MOs of the distorted molecule to those of the reference (MO overlap matrix before and after alignment). If the alignment and displacement criteria are satisfied, SQD is carried out with the reference bitstrings (red path); otherwise a new reference is defined and re-sampled on the QPU (blue path).}
\label{fig:schematic}
\end{figure}

With all key components of the VGSQD algorithm introduced, we describe the full algorithm below. It can be applied to any problem where SQD is used to determine the ground state of closely related molecular geometries, such as geometry optimization, nudged elastic band (NEB) calculations and molecular dynamics (MD). A schematic of the algorithm is shown in Figure~\ref{fig:schematic}.  The VGSQD algorithm is defined by:
\begin{enumerate}[label=(\roman*)]
\item A Hartree-Fock calculation defines the MO basis of the reference molecule $A$.
\item A set of bitstrings $\tilde{\chi}$ is sampled from a LUCJ circuit representing the reference molecule. The ground state of the reference molecule is determined with the configuration recovery and subspace diagonalization procedure of SQD and the determinants $\{\ket{x_i}\}$ defining the corresponding subspace $\mathcal{S}$ are stored. VGSQD terminates if the optimization/NEB/MD has converged.
\item The optimization/NEB/MD algorithm updates the nuclear coordinates, which define the new molecule $B$. A Hartree-Fock calculation defines the MO basis of molecule $B$.
\item The maximally aligned rotated MO basis $\tilde{B}$ is determined and  $S_{\mathrm{MO}}^{(\tilde{B}A)}$ is compared to the identity. If any component $\delta_{\mathrm{align}} = \max\big( |S_{\mathrm{MO}}^{(\tilde{B}A)} - I|_{pq} \big)$ exceeds a threshold $\epsilon_{\mathrm{align}}$ (typically 0.03), the electronic structure has changed too much for the stored samples to remain representative. $\tilde{B}$ becomes the new reference and return to step (ii). Otherwise go to (v).
\item Compute the displacement between the two molecules $\Delta\mathbf{R}=\mathbf{R}_{B}-\mathbf{R}_A$. If no component of $\lvert\Delta\mathbf{R}\rvert$  exceeds a threshold $\epsilon_{\mathrm{pos}}$ (typically 0.02 Å), go to step (vi). Otherwise, the ground state of $\tilde{B}$ is determined with configuration-recovery and subspace diagonalization using the reference bitstrings $\tilde{\chi}$. This step generates a new subspace $\mathcal{S}$ but still requires no quantum circuit execution. VGSQD terminates if the optimization/NEB/MD has converged; otherwise go to step (iii).
\item The rotated Hamiltonian is projected in $\mathcal{S}$ and diagonalized. VGSQD terminates if the optimization/NEB/MD has converged; otherwise go to step (iii).
\end{enumerate}

The agreement of the molecular orbital bases between the reference and distorted molecules allows bitstrings reuse and is a useful criterion to screen out several electronic transitions. Nevertheless, it can miss electronic transitions of the many-body ground state. For this reason, resampling on the QPU (step (ii)) is also performed when $\delta_{\mathrm{CI}} = 1 - \big|\mathbf{c}^{(\mathbf{A})\dagger} \mathbf{c}^{(\tilde{\mathbf{B}})}\big|^2 > \epsilon_{\mathrm{CI}}$, with $\mathbf{c}^{\mathbf{(A)}}$ and $\mathbf{c}^{(\mathbf{\tilde{B}})}$ the ground states' CI coefficients of the two molecules. This equals the infidelity between the reference and distorted molecules' ground states when the two MO bases are identical. We use $\epsilon_{\mathrm{CI}} = 0.01$.

For molecular geometry optimization, we use a trust-region version of the BFGS algorithm\cite{head1985,nocedal2006}. Convergence is achieved when the magnitude of all components of the forces is below 0.1~mHa/Bohr and the per-iteration energy reduction is below 1~\textmu Ha. For CI-NEB, a convergence threshold of 1 mHa/Bohr was used.

\subsection*{Details of the QPU execution}
All our experiments were run on the \texttt{ibm-quebec} Heron \texttt{r2} device. For each LUCJ circuit, transpiled using \texttt{optimization\_level=3} of the Qiskit transpiler, we carry out 256\,000 shots. Given that this is a sampling-based problem, the options available for error suppression in Qiskit IBM Runtime (\texttt{v0.43.1}) are measurement twirling, Pauli twirling, and dynamical decoupling \cite{ibm_qiskit_runtime}. To
select the best combination, we run a mirrored circuit — our circuit composed with its inverse — and measure single-Z observables on all active qubits. We test all combinations of the aforementioned error suppression strategies (including with all of them turned off),
and then select the combination with mean expectation value
across all observables closest to the ideal value of 1. This leads us to use only measurement twirling (enabled with the default settings in Qiskit IBM Runtime) in our experiments.

Summaries of circuit parameters and QPU performance are presented in Table~\ref{tab:circuits} and Table~\ref{tab:qpu}. We then compute the ground state of the reference molecule using the subspace diagonalization and the self-consistent configuration recovery procedure detailed in ref.~\citenum{robledo2025} and implemented using \texttt{qiskit\_addon\_sqd} \cite{qiskit-addon-sqd} with 5 batches of up to 9000 samples at each iteration, leading to final subspace sizes of up to 9.6E7. Classical quantum chemistry computations are done using PySCF, and the diagonalization is done using DICE\cite{holmes2016,sharma2017} on 378 vCPUs distributed across 14 compute nodes with 230 GB of memory per node of the Harfang cluster.

\begin{table}[htbp]
\centering
\caption{Circuit parameters for selected simulations presented in this work. The Be$_2$ rows correspond to simulations shown in Figure~\ref{fig:be2}; the H$_2$O rows correspond to the simulations shown in Table~\ref{tab:results} and the H$_3$ row corresponds to the first optimized image of the simulation shown in Figure~\ref{fig:neb}.}
\label{tab:circuits}
\begin{tabular}{lrrrr}
\toprule
 & Number of qubits & Circuit depth & 2Q depth & 2Q count \\
\midrule
Be$_2$ -- cc-pVDZ & 62 & 873& 198& 4586\\
Be$_2$ -- cc-pVTZ & 124 & 2033& 477& 21370\\
H$_2$O -- cc-pVDZ & 53& 826& 194& 3394\\
H$_2$O -- cc-pVTZ & 122 & 2284& 561& 19634\\
H$_3$ -- cc-pVTZ & 85& 2979& 637& 19742\\
\bottomrule
\end{tabular}
\end{table}
\begin{table}[htbp]
\centering
\caption{QPU performance metrics recorded from the hardware at the time of each of the calculations carried out in this work. The Be$_2$ rows correspond to simulations shown in Figure~\ref{fig:be2}; the H$_2$O rows correspond to the simulations shown in Table~\ref{tab:results} and the H$_3$ row corresponds to the first optimized image of the simulation shown in Figure~\ref{fig:neb}.  Median values $\pm$  standard deviation are reported for all qubits of the circuits for gates, and for all qubits excluding ancillas for readout. }
\label{tab:qpu}
\small
\begin{tabular}{l>{\centering\arraybackslash}p{1.5cm}>{\centering\arraybackslash}p{1.9cm}>{\centering\arraybackslash}p{1.5cm}>{\centering\arraybackslash}p{1.9cm}>{\centering\arraybackslash}p{1.5cm}>{\centering\arraybackslash}p{1.9cm}}
\toprule
 & 1Q gate time (ns) & 1Q gate error (\%) & 2Q gate time (ns) & 2Q gate error (\%) & Readout time (ns) & Readout error (\%) \\
\midrule
Be$_2$ -- cc-pVDZ & 32 & $0.012 \pm 0.002$& 68 & $0.166 \pm 0.081$& 2180& $0.513 \pm 0.214$\\
Be$_2$ -- cc-pVTZ & 32 & $0.017 \pm 0.004$& 68 & $0.178 \pm 0.080$& 2180& $0.610 \pm 0.317$\\
H$_2$O -- cc-pVDZ & 32 & $0.015 \pm 0.004$& 68 & $0.194 \pm 0.079$& 2180& $0.476 \pm 0.183$\\
H$_2$O -- cc-pVTZ & 32 & $0.019 \pm 0.005$& 68 & $0.160 \pm 0.059$& 2584& $0.562 \pm 0.220$\\
H$_3$ -- cc-pVTZ & 32 & $0.017 \pm 0.005$& 68 & $0.205 \pm 0.087$& 2180 & $0.562 \pm 0.244$\\
\bottomrule
\end{tabular}
\end{table}

\section*{Results and discussion}

\begin{figure}[htbp]
\centering
\includegraphics[width=\linewidth]{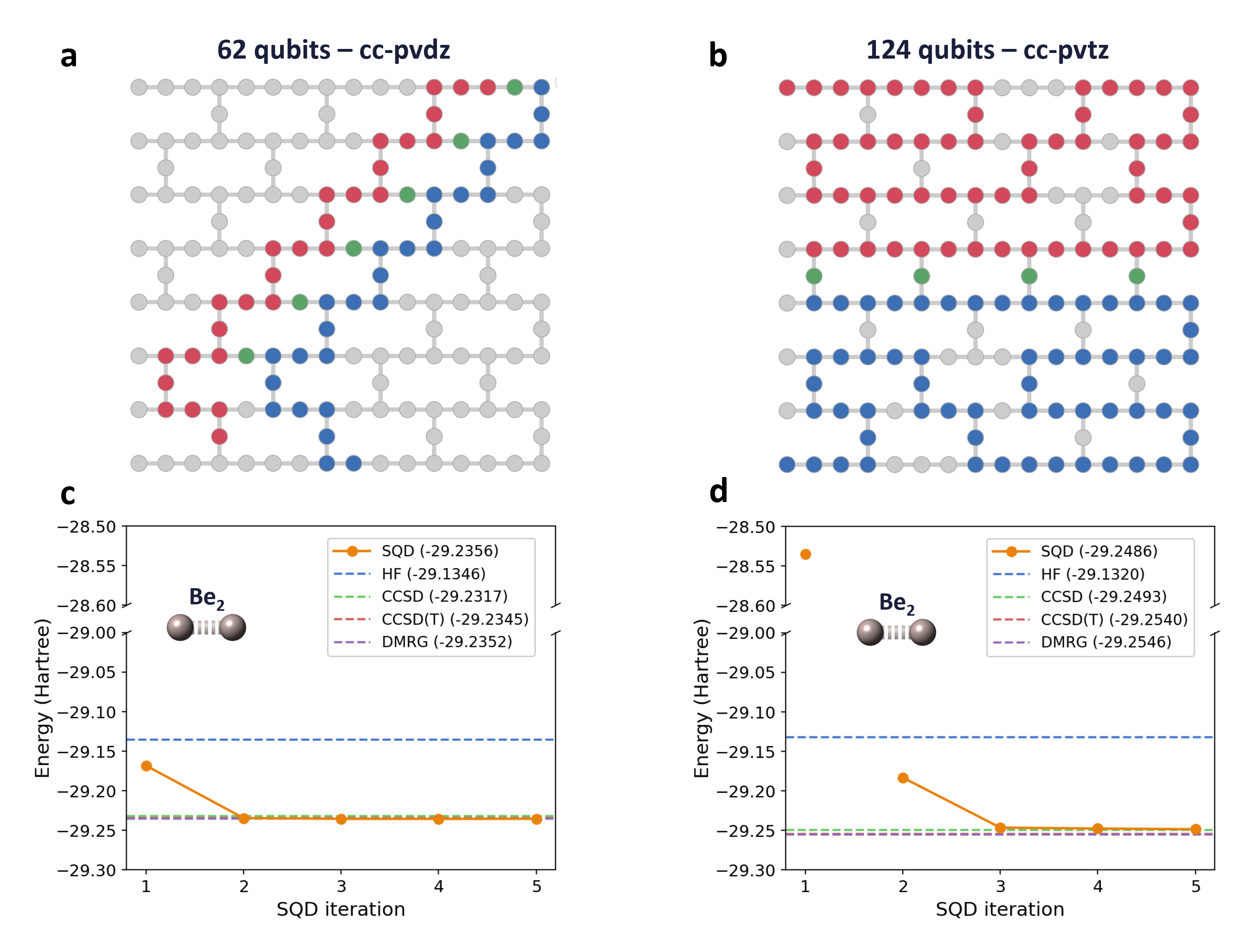}
\caption{(a, b) Qubit layouts of LUCJ circuits of Be$_2$ with the cc-pVDZ (a) and cc-pVTZ (b) basis sets. Qubits used to encode the occupation numbers of $\alpha$ ($\beta$) spin orbitals are shown in red (blue). Ancilla qubits mediating the density--density interactions between opposite spin are shown in green. (c, d) SQD energy convergence vs.\ iteration of the configuration recovery procedure for Be$_2$ with the cc-pVDZ basis set at bond length of 2.564 Å (c) and cc-pVTZ basis set at bond length of 2.385 Å (d), with comparison to classical methods.}
\label{fig:be2}
\end{figure}

We first examine a simple, yet challenging use case, which is the beryllium dimer (Be$_2$)\cite{merritt2009}. Its weak covalent bond is driven by a delicate balance of static and dynamic correlation, making accurate bond length and energy predictions difficult for conventional classical methods. Before carrying out molecular geometry optimization with VGSQD, we validate the SQD ground state calculation and benchmark against classical methods using two basis sets: cc-pVDZ and cc-pVTZ. The cc-pVDZ basis set leads to a 62-qubit circuit (Figure~\ref{fig:be2}a), whereas 124 qubits are used for cc-pVTZ (Figure~\ref{fig:be2}b).

The convergence of the SQD ground state energies during the configuration recovery procedure is reported in Figure~\ref{fig:be2}c and Figure~\ref{fig:be2}d. For both basis sets, convergence is achieved within three iterations, followed by minor statistical fluctuations. The converged SQD ground state energies are $-29.2356$~Ha and $-29.2486$~Ha for cc-pVDZ and cc-pVTZ respectively. SQD is on par with CCSD for the larger basis set and outperforms CCSD(T)/DMRG \footnote{See Appendix~A for the details of the DMRG calculations.} for the smaller basis set, thus validating the accuracy of SQD for the Be$_2$ molecule. We note that SQD slightly underperforms in comparison with CCSD(T) and DMRG for the larger, cc-pVTZ basis set. This is likely due to the corresponding quantum circuits being deeper and due to the increased memory requirements for classical postprocessing, which limited the size of the subspace onto which the Hamiltonian was projected. Nevertheless, these results suggest that Be$_2$ with the cc-pVDZ basis set is a strong use case to test the VGSQD algorithm.

\begin{figure}[htbp]
\centering
\includegraphics[width=\linewidth]{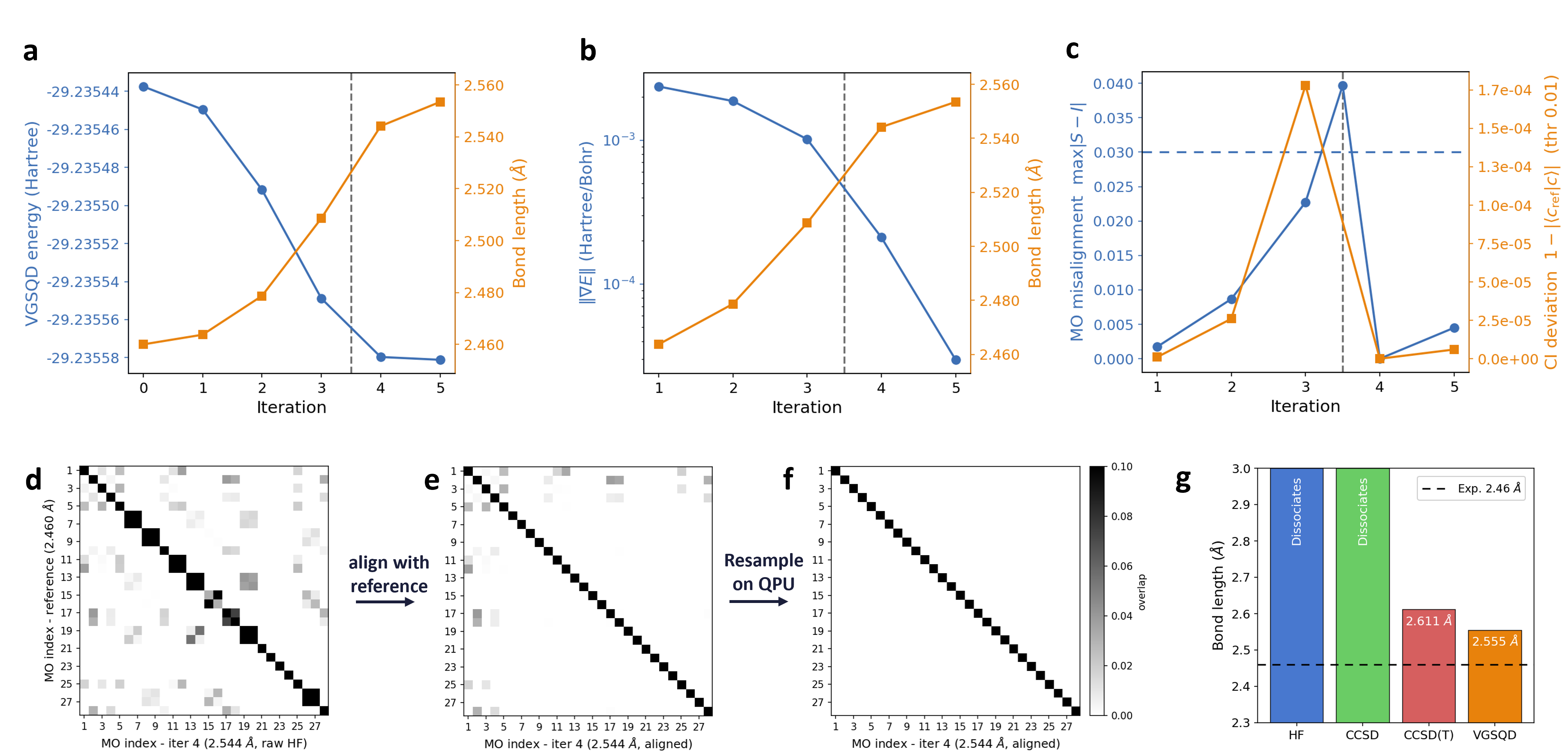}
\caption{VGSQD geometry optimization of Be$_2$ with the cc-pVDZ basis set. (a, b) SQD energy (a) and gradient norm (b) vs.\ iteration, with the bond length shown on the right axis. (c) MO misalignment $\delta_{\mathrm{align}}$ and CI infidelity $\delta_{\mathrm{CI}}$ vs.\ iteration; the horizontal dashed line indicates $\epsilon_{\mathrm{align}}$. The vertical dashed lines in (a--c) mark the re-sampling on the QPU. (d--f) Cross-geometry MO overlap matrix between the reference (2.460~Å) and iteration 4 (2.544~Å) geometries before alignment (d), after alignment (e) and after re-sampling on the QPU (f). (g) Optimized bond length with comparison to classical methods and experiment.}
\label{fig:be2opt}
\end{figure}

Next, we examine the application of the VGSQD algorithm to molecular geometry optimization for the beryllium dimer with the cc-pVDZ basis set. We start the optimization from a bond length of 2.46~Å (experimental value). The energy and gradient norm are shown in Figure~\ref{fig:be2opt}a and Figure~\ref{fig:be2opt}b, respectively. Both quantities decrease monotonically and the optimization terminates successfully when the energy and gradient criteria are fulfilled at iteration 5. We note that each self-consistent configuration recovery and diagonalization yield a slightly different subspace. While the ground state solution in the subspace is a good approximation of the FCI solution, it only converges to the exact solution in the FCI limit. While not observed here, it can cause the energy and gradient to increase during the optimization due to a discontinuous potential energy surface\cite{motta2023}. Care must thus be taken to ensure convergence of the SQD energy \emph{with respect to} the subspace dimension.

The MO and CI criteria during the optimization are plotted in Figure~\ref{fig:be2opt}c. As expected, both the infidelity $\delta_{\mathrm{CI}}$ and the MO misalignment $\delta_{\mathrm{align}}$ increase between QPU sampling rounds, during the Be$_2$ molecular geometry optimization. Here, only one re-sampling on the QPU was required, as indicated by the dashed line between iteration 3 and 4 and triggered by the $\delta_{\mathrm{align}}$ criterion. Evidently, both $\delta_{\mathrm{CI}}$ and $\delta_{\mathrm{align}}$ return to zero after updating the reference and resampling on the QPU, then start slowly increasing again. We note that $\delta_{\mathrm{align}}$ increases much more rapidly than $\delta_{\mathrm{CI}}$, which is expected considering the translation of the nuclei. Interestingly, $\delta_{\mathrm{CI}}$ remains very small and considerably below $10^{-3}$. The agreement of the ground state CI coefficients from the two geometries confirms that the MO rotation makes the sampled bitstrings re-usable across geometries and suggests that the VGSQD algorithm converges to the ground state, even when using bitstrings sampled from another molecular geometry. Nonetheless, these criteria may not be sufficient to identify problematic electronic transitions that make the sampled bitstrings no longer re-usable in molecules with near-degenerate ground states. In those cases, the VGSQD algorithm may need to be modified to track multiple states, including low-lying excited states, such as with the ExtSQD algorithm\cite{barison2025}.

The results of the MO alignment procedure are shown in Figure~\ref{fig:be2opt}(d--f). The unaligned cross-geometry MO overlap matrix $S_{\mathrm{MO}}^{(AB)}$ at iteration 4, where resampling on the QPU was triggered by the $\delta_{\mathrm{align}}$ criterion, is shown in Figure~\ref{fig:be2opt}d. There are large off-diagonal elements, which are mainly attributed to arbitrary rotations of the MOs within the occupied and virtual Hartree-Fock subspaces. These rotations alter the correspondence between MOs at the two geometries, making the sampled bitstrings no longer correspond to the same orbital configurations at the new geometry. The MO overlap matrix after alignment with the reference molecule's MO basis $S_{\mathrm{MO}}^{(A\tilde{B})}$ can be seen in Figure~\ref{fig:be2opt}e. $S_{\mathrm{MO}}^{(A\tilde{B})}$ is mostly diagonal with some small off-diagonal elements (magnified by the compressed color scale range), thus indicating that correspondence between MOs at the two geometries has been restored. Here, the remaining small off-diagonal elements are expected since the MOs span slightly different subspaces because of the changes in the nuclear coordinates and resulting AO bases. As the reference molecule is updated to the new geometry and its LUCJ circuit is sampled on the QPU, $S_{\mathrm{MO}}^{(A\tilde{B})}$ returns to the identity (Figure~\ref{fig:be2opt}f). The resulting fresh bitstrings can then be used for the subsequent geometry optimization iterations, while satisfying the $\delta_{\mathrm{align}}$ and $\delta_{\mathrm{CI}}$ tolerance.

The VGSQD geometry optimization results for Be$_2$ with the cc-pVDZ basis are displayed in Figure~\ref{fig:be2opt}g and compared to classical methods. The geometry optimization yields unbounded monomers, rather than a weakly bonded dimer for both Hartree-Fock and CCSD. On the other hand, CCSD(T) and VGSQD recover a bound dimer as they both incorporate higher-order excitations required to describe the correlation. Interestingly, the bond length found with VGSQD (2.555~Å) is closer to the experimental value (2.46~Å) than the CCSD(T) one (2.611~Å). This improvement over CCSD(T) is consistent with the energy comparison reported in the previous section. These results support the validity of the VGSQD algorithm for molecular geometry optimization in classically challenging strongly correlated systems.

While optimizing the Be$_2$ bond length is a useful case study to validate the VGSQD algorithm, Be$_2$ only has a single degree of freedom, which makes it approachable by carrying out SQD calculations at various bond lengths to find the equilibrium interatomic distance. Nonetheless, this approach has lower accuracy and a higher quantum and classical computational cost. Furthermore, it scales exponentially with the number of degrees of freedom (DOFs). On the other hand, VGSQD can \emph{a priori} be applied to molecules with large numbers of DOFs. Although there are no theoretical scaling laws for geometry optimization, a linear scaling of the number of iterations with the number of DOFs is generally observed\cite{wang2016}. Furthermore, it is reasonable to expect better scaling for VGSQD since QPU calls are needed mainly when the geometry changes significantly, which can largely be avoided by sensible initialization.

In Table~\ref{tab:results} below, we also demonstrate the application of VGSQD to H$_2$O geometry optimization\footnote{Additional convergence results are presented in Appendix~B.}. With the cc-pVTZ basis set, the bond length and bond angle converge to the experimental value within 0.002~Å and 0.41\textdegree, respectively. This confirms the potential of VGSQD for geometry optimization of molecules with multiple DOFs. Nonetheless, we note that CCSD(T) outperforms VGSQD on the energy prediction for both the cc-pVDZ and cc-pVTZ basis sets, as expected given that the correlation in H$_2$O is mainly dynamical, which is well captured by CCSD(T).

\begin{table}[htbp]
\centering
\caption{VGSQD energies, bond lengths and bond angles with comparison to CCSD(T) and experimental values. Energies are reported from the corresponding equilibrium geometry obtained for each method using BFGS. Experimental energies are not included since absolute electronic ground-state energies are not directly accessible experimentally. Experimental Be$_2$ bond angle is omitted as it is a diatomic molecule. }
\label{tab:results}
\begin{tabular}{lccc}
\toprule
 & Energy (Ha) & Bond length (Å) & Bond angle (\textdegree) \\
\midrule
Be$_2$ experimental & --- & 2.460 & --- \\
Be$_2$ VGSQD -- cc-pVDZ & $-29.2356$& 2.555& --- \\
Be$_2$ VGSQD -- cc-pVTZ & $-29.2485$& 2.383& --- \\
Be$_2$ CCSD(T) -- cc-pVDZ & $-29.2346$& 2.611 & --- \\
Be$_2$ CCSD(T) -- cc-pVTZ & $-29.2541$& 2.340 & --- \\
\midrule
H$_2$O experimental & --- & 0.958 & 104.48 \\
H$_2$O VGSQD -- cc-pVDZ & $-76.2417$& 0.966 & 102.16 \\
H$_2$O VGSQD -- cc-pVTZ & $-76.3021$& 0.956 & 104.07 \\
H$_2$O CCSD(T) -- cc-pVDZ & $-76.2434$ & 0.966 & 101.94\\
H$_2$O CCSD(T) -- cc-pVTZ & $-76.3458$ & 0.958 & 103.68 \\
\bottomrule
\end{tabular}
\end{table}

\begin{figure}[htbp]
\centering
\includegraphics[width=\linewidth]{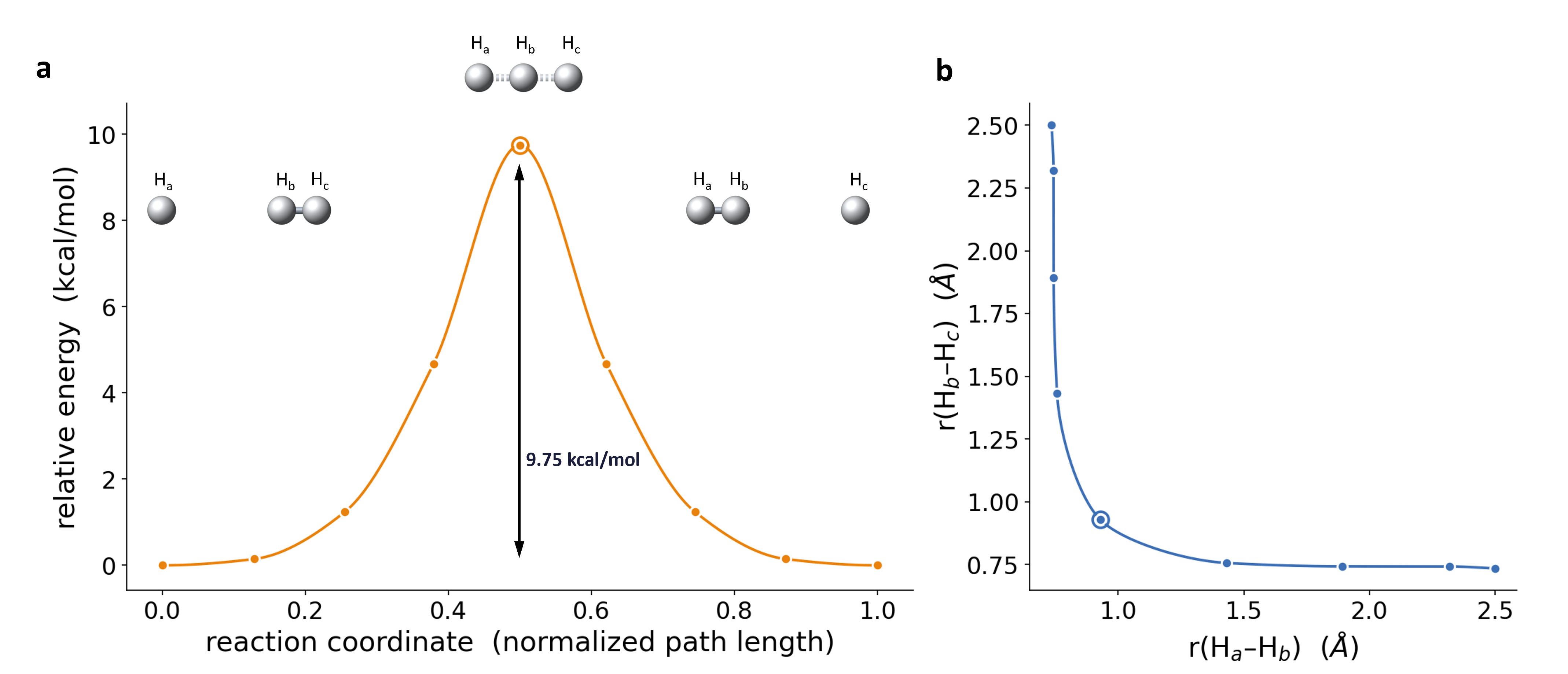}
\caption{Minimum energy path of the $\mathrm{H_2} + \mathrm{H} \rightarrow \mathrm{H} + \mathrm{H_2}$ reaction obtained with VGSQD and CI-NEB (cc-pVTZ). (a) Energy of the NEB images relative to the reactants vs.\ normalized path length, with the transition state circled. (b) Projection of the path onto the $r(\mathrm{H}_a\text{--}\mathrm{H}_b)$ and $r(\mathrm{H}_b\text{--}\mathrm{H}_c)$ coordinates.}
\label{fig:neb}
\end{figure}

Finally, we demonstrate the application of VGSQD to chemical reactions; that is, finding their minimum energy path (MEP) and energy barrier. For this case study, we look at the $\mathrm{H_2} + \mathrm{H} \rightarrow \mathrm{H} + \mathrm{H_2}$ hydrogen exchange reaction and use the climbing image nudged elastic band method (CI-NEB) with the cc-pVTZ basis set\footnote{Additional convergence results are presented in Appendix~C.}. For each NEB image, an 85 qubit LUCJ circuit was sampled, and no resampling was triggered during the MEP optimization. While not done in this work, QPU cost could be reduced by using the same circuits for multiple images in the same, or across multiple iterations, if $\delta_{\mathrm{CI}}$ and $\delta_{\mathrm{align}}$ do not exceed their respective thresholds. Furthermore, the classical cost and memory scale almost perfectly linearly with the number of images, and images can be distributed across nodes with near-ideal memory distribution and speedup. Figure~\ref{fig:neb} shows the optimized MEP. The path is symmetric about the transition state, as expected from the symmetry of the reaction. The transition state converges to the collinear H$_3$ saddle point with an energy barrier of 9.75~kcal/mol (Figure~\ref{fig:neb}a). Projecting the path onto the $r_{\mathrm{H}_a} - r_{\mathrm{H}_b}$ and $r_{\mathrm{H}_b} - r_{\mathrm{H}_c}$ coordinates shows the characteristic L-shape profile of a concerted exchange where the incoming bond contracts, while the breaking bond extends, with a saddle point H-H distance of 0.929~Å. These results are in very good agreement with FCI calculations, which yield a 9.94~kcal/mol barrier and 0.930~Å H-H distance at the saddle point with the same basis set\cite{peterson1994}.

\section*{Conclusion}

In summary, we developed the variable geometry sample-based quantum diagonalization (VGSQD) algorithm, which extends SQD based algorithms and workflows to molecules with variable geometries by allowing quantum samples to be reused across multiple molecular configurations. We demonstrated the application of VGSQD to molecular geometry optimization by leveraging a quantum-centric supercomputing workflow using utility-scale quantum circuits of up to 124 qubits, run on an IBM superconducting gate-based quantum computer. The accuracy of the calculations rivals state-of-the-art classical methods such as CCSD(T) and DMRG, both for ground state energies and optimized molecular geometries. Moreover, we demonstrated that VGSQD can naturally be used to study chemical reactions and determine minimum energy paths, saddle points and energy barriers. These demonstrations broaden the scope of quantum algorithms in chemistry applications. While not explored in this work, VGSQD can be implemented with other ansätze than LUCJ and can be adapted to integrate with multiple other algorithms such as sample-based Krylov quantum diagonalization, TrimSQD \cite{merz2026}, and embedding methods and thus can play a key role in current and early fault tolerant quantum-centric workflows.

\section*{Acknowledgements}

The authors thank Kevin Sung, Mario Motta, and Sean Wagner for fruitful discussions and comments on the draft. The authors acknowledge access to the \texttt{ibm\_quebec} system for quantum experiments and the \textit{Harfang} cluster for classical postprocessing, through PINQ\textsuperscript{2}.
The authors used generative AI tools (Claude and GPT) to assist with writing the code used in this manuscript, to generate drafts of figures and plots, to scaffold the methods section, to brainstorm about experiments and parts of the algorithm, to help with literature review, and to help with clarity, syntax, and grammar. The authors designed and carried out the core of the experiments and algorithms and maintain full responsibility for the algorithm design, its execution and all the reported results, and conclusions.
\newpage
\bibliographystyle{unsrtnat}
\bibliography{references}

@article{kitaev1995,
  author  = {Kitaev, A. Y.},
  title   = {Quantum Measurements and the {Abelian} Stabilizer Problem},
  journal = {arXiv preprint quant-ph/9511026},
  year    = {1995}
}

@article{aspuru2005,
  author  = {Aspuru-Guzik, A. and Dutoi, A. D. and Love, P. J. and Head-Gordon, M.},
  title   = {Simulated Quantum Computation of Molecular Energies},
  journal = {Science},
  year    = {2005},
  volume  = {309},
  number  = {5741},
  pages   = {1704--1707}
}

@article{peruzzo2014,
  author  = {Peruzzo, A. and McClean, J. and Shadbolt, P. and Yung, M.-H. and Zhou, X.-Q. and Love, P. J. and Aspuru-Guzik, A. and O'Brien, J. L.},
  title   = {A Variational Eigenvalue Solver on a Photonic Quantum Processor},
  journal = {Nature Communications},
  year    = {2014},
  volume  = {5},
  number  = {1},
  pages   = {4213},
  doi     = {10.1038/ncomms5213}
}

@article{mcclean2016,
  author  = {McClean, J. R. and Romero, J. and Babbush, R. and Aspuru-Guzik, A.},
  title   = {The Theory of Variational Hybrid Quantum-Classical Algorithms},
  journal = {New Journal of Physics},
  year    = {2016},
  volume  = {18},
  number  = {2},
  pages   = {023023}
}

@article{kandala2017,
  author  = {Kandala, A. and Mezzacapo, A. and Temme, K. and Takita, M. and Brink, M. and Chow, J. M. and Gambetta, J. M.},
  title   = {Hardware-Efficient Variational Quantum Eigensolver for Small Molecules and Quantum Magnets},
  journal = {Nature},
  year    = {2017},
  volume  = {549},
  number  = {7671},
  pages   = {242--246}
}

@article{nakagawa2024,
  author  = {Nakagawa, Y. O. and Kamoshita, M. and Mizukami, W. and Sudo, S. and Ohnishi, Y.-y.},
  title   = {{ADAPT-QSCI}: Adaptive Construction of an Input State for Quantum-Selected Configuration Interaction},
  journal = {Journal of Chemical Theory and Computation},
  year    = {2024},
  volume  = {20},
  number  = {24},
  pages   = {10817--10825}
}

@article{motta2024,
  author  = {Motta, M. and Kirby, W. and Liepuoniute, I. and Sung, K. J. and Cohn, J. and Mezzacapo, A. and Klymko, K. and Nguyen, N. and Yoshioka, N. and Rice, J. E.},
  title   = {Subspace Methods for Electronic Structure Simulations on Quantum Computers},
  journal = {Electronic Structure},
  year    = {2024},
  volume  = {6},
  number  = {1},
  pages   = {013001},
  doi     = {10.1088/2516-1075/ad3592}
}

@article{robledo2025,
  author  = {Robledo-Moreno, J. and Motta, M. and Haas, H. and Javadi-Abhari, A. and Jurcevic, P. and Kirby, W. and Martiel, S. and Sharma, K. and Sharma, S. and Shirakawa, T. and Sitdikov, I. and Sun, R.-Y. and Sung, K. J. and Takita, M. and Tran, M. C. and Yunoki, S. and Mezzacapo, A.},
  title   = {Chemistry beyond the Scale of Exact Diagonalization on a Quantum-Centric Supercomputer},
  journal = {Science Advances},
  year    = {2025},
  volume  = {11},
  number  = {25},
  pages   = {eadu9991},
  doi     = {10.1126/sciadv.adu9991}
}

@article{piccinelli2025,
  author  = {Piccinelli, S. and Baiardi, A. and Rossmannek, M. and {Carrera Vazquez}, A. and Tacchino, F. and Mensa, S. and Altamura, E. and Alavi, A. and Motta, M. and Robledo-Moreno, J. and Kirby, W. and Sharma, K. and Mezzacapo, A. and Tavernelli, I.},
  title   = {Quantum Chemistry with Provable Convergence via Randomized Sample-Based Quantum Diagonalization},
  journal = {arXiv preprint arXiv:2508.02578},
  year    = {2025}
}

@article{shajan2025,
  author  = {Shajan, A. and Kaliakin, D. and Mitra, A. and {Robledo Moreno}, J. and Li, Z. and Motta, M. and Johnson, C. and Saki, A. A. and Das, S. and Sitdikov, I. and Mezzacapo, A. and Merz, Jr., K. M.},
  title   = {Toward Quantum-Centric Simulations of Extended Molecules: Sample-Based Quantum Diagonalization Enhanced with Density Matrix Embedding Theory},
  journal = {Journal of Chemical Theory and Computation},
  year    = {2025},
  volume  = {21},
  number  = {14},
  pages   = {6801--6810},
  doi     = {10.1021/acs.jctc.5c00114}
}

@article{yu2025,
  author  = {Yu, J. and {Robledo Moreno}, J. and Iosue, J. T. and Bertels, L. and Claudino, D. and Fuller, B. and Groszkowski, P. and Humble, T. S. and Jurcevic, P. and Kirby, W. and Maier, T. A. and Motta, M. and Pokharel, B. and Seif, A. and Shehata, A. and Sung, K. J. and Tran, M. C. and Tripathi, V. and Mezzacapo, A. and Sharma, K.},
  title   = {Quantum-Centric Algorithm for Sample-Based {Krylov} Diagonalization},
  journal = {arXiv preprint arXiv:2501.09702},
  year    = {2025}
}

@article{piccinelli2026,
  author  = {Piccinelli, S. and Barison, S. and Baiardi, A. and Tacchino, F. and Repp, J. and Ron{\v{c}}evi{\'c}, I. and Albrecht, F. and Anderson, H. L. and Gross, L. and Curioni, A. and Tavernelli, I.},
  title   = {Exploring Pathways towards Quantum Advantage in Quantum Chemistry: The Case of a Molecule with Half-{M{\"o}bius} Topology},
  journal = {arXiv preprint arXiv:2603.08696},
  year    = {2026}
}

@article{merz2026,
  author  = {Merz, Jr., K. M. and Shajan, A. and Kaliakin, D. and Liang, F. and Otsuka, Y. and Shirakawa, T. and Broers, L. and Xu, H. and Tsuji, M. and Sato, M. and Yunoki, S. and Wakizaka, R. and Kawashima, Y. and Doi, J. and Takahashi, H. and Itoko, T. and Horii, H. and Pellegrini, T. and {Robledo Moreno}, J. and Sung, K. J. and Fejer, E. and Walkup, R. and Seelam, S. and Motta, M.},
  title   = {Crossing the 12,000-Atom Barrier with Heterogeneous Quantum-Classical Supercomputing: Quantum Chemistry of Protein-Ligand Complexes},
  journal = {arXiv preprint arXiv:2605.01138},
  year    = {2026}
}

@article{merz2025,
  author  = {Kaliakin, D. and Shajan, A. and Liang, F. and {Robledo Moreno}, J. and Li, Z. and Mitra, A. and Motta, M. and Johnson, C. and Saki, A. A. and Das, S. and Sitdikov, I. and Mezzacapo, A. and Merz, Jr., K. M.},
  title   = {Accurate Quantum-Centric Simulations of Intermolecular Interactions},
  journal = {Communications Physics},
  year    = {2025},
  volume  = {8},
  number  = {1},
  pages   = {396}
}

@article{kaliakin2025,
  author  = {Kaliakin, D. and Shajan, A. and Liang, F. and Merz, Jr., K. M.},
  title   = {Implicit Solvent Sample-Based Quantum Diagonalization},
  journal = {The Journal of Physical Chemistry B},
  year    = {2025},
  volume  = {129},
  number  = {23},
  pages   = {5788--5796}
}

@article{barison2025,
  author  = {Barison, S. and {Robledo Moreno}, J. and Motta, M.},
  title   = {Quantum-Centric Computation of Molecular Excited States with Extended Sample-Based Quantum Diagonalization},
  journal = {Quantum Science and Technology},
  year    = {2025},
  volume  = {10},
  number  = {2},
  pages   = {025034},
  doi     = {10.1088/2058-9565/adb781}
}

@article{liepuoniute2025,
  author  = {Liepuoniute, I. and Doney, K. D. and {Robledo Moreno}, J. and Job, J. A. and Friend, W. S. and Jones, G. O.},
  title   = {Quantum-Centric Computational Study of Methylene Singlet and Triplet States},
  journal = {Journal of Chemical Theory and Computation},
  year    = {2025},
  volume  = {21},
  number  = {10},
  pages   = {5062--5070},
  doi     = {10.1021/acs.jctc.5c00075}
}

@article{smith2025,
  author  = {Smith, T. and Gujarati, T. P. and Motta, M. and Link, B. and Liepuoniute, I. and Friedhoff, T. and Nishimura, H. and Nguyen, N. and Williams, K. S. and {Robledo Moreno}, J. and Johnson, C. and Sung, K. J. and Saki, A. A. and Kagele, M.},
  title   = {Quantum-Centric Simulation of Hydrogen Abstraction by Sample-Based Quantum Diagonalization and Entanglement Forging},
  journal = {arXiv preprint arXiv:2508.08229},
  year    = {2025}
}

@article{shivpuje2025,
  author  = {Shivpuje, S. and Gujarati, T. P. and Van, R. and Pickard, IV, F. C. and Friedhoff, T. and Liepuoniute, I. and Davis, W. and Jones, G. O. and Galda, A.},
  title   = {Sample-Based Quantum Diagonalization Methods for Modeling the Photochemistry of Diazirine and Diazo Compounds},
  journal = {arXiv preprint arXiv:2510.00484},
  year    = {2025}
}

@article{yamamoto2026,
  author  = {Yamamoto, K. and Masui, R. and Nakajima, T. and Tsuji, M. and Sato, M. and Schow, P. and Heidemann, L. and Burke, M. and Seitz, P. and Backhouse, O. J. and Pedersen, J. W. and Children, J. and Holliman, C. and Lysne, N. and Okuno, D. and Sivarajah, S. and {Mu{\~n}oz Ramo}, D. and Chernoguzov, A. and Duncan, R.},
  title   = {Quantum-{HPC} Hybrid Computation of Biomolecular Excited-State Energies},
  journal = {arXiv preprint arXiv:2601.15677},
  year    = {2026}
}

@incollection{jonsson1998,
  author    = {J{\'o}nsson, H. and Mills, G. and Jacobsen, K. W.},
  title     = {Nudged Elastic Band Method for Finding Minimum Energy Paths of Transitions},
  booktitle = {Classical and Quantum Dynamics in Condensed Phase Simulations},
  publisher = {World Scientific},
  year      = {1998},
  pages     = {385--404}
}

@article{henkelman2000a,
  author  = {Henkelman, G. and Uberuaga, B. P. and J{\'o}nsson, H.},
  title   = {A Climbing Image Nudged Elastic Band Method for Finding Saddle Points and Minimum Energy Paths},
  journal = {The Journal of Chemical Physics},
  year    = {2000},
  volume  = {113},
  number  = {22},
  pages   = {9901--9904}
}

@article{henkelman2000b,
  author  = {Henkelman, G. and J{\'o}nsson, H.},
  title   = {Improved Tangent Estimate in the Nudged Elastic Band Method for Finding Minimum Energy Paths and Saddle Points},
  journal = {The Journal of Chemical Physics},
  year    = {2000},
  volume  = {113},
  number  = {22},
  pages   = {9978--9985}
}

@article{sun2018,
  author  = {Sun, Q. and Berkelbach, T. C. and Blunt, N. S. and Booth, G. H. and Guo, S. and Li, Z. and Liu, J. and McClain, J. D. and Sayfutyarova, E. R. and Sharma, S. and Wouters, S. and Chan, G. K.-L.},
  title   = {{PySCF}: The {Python}-based Simulations of Chemistry Framework},
  journal = {Wiley Interdisciplinary Reviews: Computational Molecular Science},
  year    = {2018},
  volume  = {8},
  number  = {1},
  pages   = {e1340}
}

@article{jordan1928,
  author  = {Jordan, P. and Wigner, E.},
  title   = {{\"U}ber das {Paulische} {\"A}quivalenzverbot},
  journal = {Zeitschrift f{\"u}r Physik},
  year    = {1928},
  volume  = {47},
  number  = {9},
  pages   = {631--651}
}

@article{neuscamman2013,
  author  = {Neuscamman, E.},
  title   = {Communication: A {Jastrow} Factor Coupled Cluster Theory for Weak and Strong Electron Correlation},
  journal = {The Journal of Chemical Physics},
  year    = {2013},
  volume  = {139},
  number  = {18},
  pages   = {181101}
}

@article{matsuzawa2020,
  author  = {Matsuzawa, Y. and Kurashige, Y.},
  title   = {{Jastrow}-Type Decomposition in Quantum Chemistry for Low-Depth Quantum Circuits},
  journal = {Journal of Chemical Theory and Computation},
  year    = {2020},
  volume  = {16},
  number  = {2},
  pages   = {944--952}
}

@article{motta2023,
  author  = {Motta, M. and Sung, K. J. and Whaley, K. B. and Head-Gordon, M. and Shee, J.},
  title   = {Bridging Physical Intuition and Hardware Efficiency for Correlated Electronic States: The Local Unitary Cluster {Jastrow} Ansatz for Electronic Structure},
  journal = {Chemical Science},
  year    = {2023},
  volume  = {14},
  number  = {40},
  pages   = {11213--11227}
}

@article{holmes2016,
  author  = {Holmes, A. A. and Tubman, N. M. and Umrigar, C. J.},
  title   = {Heat-Bath Configuration Interaction: An Efficient Selected Configuration Interaction Algorithm Inspired by Heat-Bath Sampling},
  journal = {Journal of Chemical Theory and Computation},
  year    = {2016},
  volume  = {12},
  number  = {8},
  pages   = {3674--3680}
}

@article{sharma2017,
  author  = {Sharma, S. and Holmes, A. A. and Jeanmairet, G. and Alavi, A. and Umrigar, C. J.},
  title   = {Semistochastic Heat-Bath Configuration Interaction Method: Selected Configuration Interaction with Semistochastic Perturbation Theory},
  journal = {Journal of Chemical Theory and Computation},
  year    = {2017},
  volume  = {13},
  number  = {4},
  pages   = {1595--1604}
}

@article{king1967,
  author  = {King, H. F. and Stanton, R. E. and Kim, H. and Wyatt, R. E. and Parr, R. G.},
  title   = {Corresponding Orbitals and the Nonorthogonality Problem in Molecular Quantum Mechanics},
  journal = {The Journal of Chemical Physics},
  year    = {1967},
  volume  = {47},
  number  = {6},
  pages   = {1936--1941}
}

@article{olsen1995,
  author  = {Olsen, J. and Bak, K. L. and Ruud, K. and Helgaker, T. and J{\o}rgensen, P.},
  title   = {Orbital Connections for Perturbation-Dependent Basis Sets},
  journal = {Theoretica Chimica Acta},
  year    = {1995},
  volume  = {90},
  number  = {5},
  pages   = {421--439}
}

@article{coe2022,
  author  = {Coe, J. P. and {Moreno Carrascosa}, A. and Simmermacher, M. and Kirrander, A. and Paterson, M. J.},
  title   = {Efficient Computation of Two-Electron Reduced Density Matrices via Selected Configuration Interaction},
  journal = {Journal of Chemical Theory and Computation},
  year    = {2022},
  volume  = {18},
  number  = {11},
  pages   = {6690--6699}
}

@article{coe2023,
  author  = {Coe, J. P.},
  title   = {Analytic Gradients for Selected Configuration Interaction},
  journal = {Journal of Chemical Theory and Computation},
  year    = {2023},
  volume  = {19},
  number  = {3},
  pages   = {874--886}
}

@article{head1985,
  author  = {Head, J. D. and Zerner, M. C.},
  title   = {A {Broyden--Fletcher--Goldfarb--Shanno} Optimization Procedure for Molecular Geometries},
  journal = {Chemical Physics Letters},
  year    = {1985},
  volume  = {122},
  number  = {3},
  pages   = {264--270}
}

@book{nocedal2006,
  author    = {Nocedal, J. and Wright, S.},
  title     = {Numerical Optimization},
  publisher = {Springer},
  year      = {2006}
}

@article{merritt2009,
  author  = {Merritt, J. M. and Bondybey, V. E. and Heaven, M. C.},
  title   = {Beryllium Dimer---Caught in the Act of Bonding},
  journal = {Science},
  year    = {2009},
  volume  = {324},
  number  = {5934},
  pages   = {1548--1551}
}

@article{wang2016,
  author  = {Wang, L.-P. and Song, C.},
  title   = {Geometry Optimization Made Simple with Translation and Rotation Coordinates},
  journal = {The Journal of Chemical Physics},
  year    = {2016},
  volume  = {144},
  number  = {21},
  pages   = {214108}
}

@article{peterson1994,
  author  = {Peterson, K. A. and Woon, D. E. and Dunning, Jr., T. H.},
  title   = {Benchmark Calculations with Correlated Molecular Wave Functions. {IV}. The Classical Barrier Height of the $\mathrm{H} + \mathrm{H_2} \rightarrow \mathrm{H_2} + \mathrm{H}$ Reaction},
  journal = {The Journal of Chemical Physics},
  year    = {1994},
  volume  = {100},
  number  = {10},
  pages   = {7410--7415}
}

@misc{ibm_qiskit_runtime,
  author       = {{IBM Quantum}},
  title        = {IBM Quantum Compute (formerly Qiskit Runtime) Documentation},
  year         = {2026},
  howpublished = {\url{https://quantum.cloud.ibm.com/docs/en/api/qiskit-ibm-runtime}},
  note         = {Accessed: 2026-10-02}
}

@article{pyblock2,
  author  = {Zhai, H. and Larsson, H. R. and Lee, S. and Cui, Z. and Zhu, T. and Sun, C. and Peng, L. and Peng, R. and Liao, K. and Tölle, J. and Yang, J. and Li, S. and Chan, G. K.-L.},
  title   = {Block2: a comprehensive open source framework to develop and apply state-of-the-art DMRG algorithms in electronic structure and beyond},
  journal = {The Journal of Chemical Physics},
  year    = {2023},
  volume  = {159},
  pages   = {234801},
  doi     = {10.1063/5.0180424}
}

@article{filippi2000correlated,
  title={Correlated sampling in quantum Monte Carlo: A route to forces},
  author={Filippi, Claudia and Umrigar, CJ},
  journal={Physical Review B},
  volume={61},
  number={24},
  pages={R16291},
  year={2000},
  publisher={APS}
}

@article{sung2026ffsim,
      title={ffsim: Faster simulation of fermionic quantum circuits},
      author={Kevin J. Sung and Inho Choi and Mirko Amico and Bartholomew Andrews and Esra Ayantuna and Yukio Kawashima and Wan-Hsuan Lin and David Omanovic and Samuele Piccinelli and Javier Robledo Moreno and Abdullah Ash Saki and James Shee and Soyoung Shin and Minh C. Tran and Kento Ueda and Haimeng Zhang and Mario Motta},
      journal={arXiv preprint arXiv:2605.03123},
      year={2026}
}

@misc{qiskit-addon-sqd,
  author = {
    Abdullah Ash Saki
    and Stefano Barison
    and Bryce Fuller
    and James R. Garrison
    and Jennifer R. Glick
    and Caleb Johnson
    and Antonio Mezzacapo
    and Javier Robledo-Moreno
    and Max Rossmannek
    and Paul Schweigert
    and Iskandar Sitdikov
    and Kevin J. Sung
  },
  title = {{Qiskit addon: sample-based quantum diagonalization}},
  howpublished = {\url{https://github.com/Qiskit/qiskit-addon-sqd}},
  year = {2024}
}

@article{dmn4-snfx,
  title = {Quantum-selected configuration interaction: Classical diagonalization of Hamiltonians in subspaces selected by quantum computers},
  author = {Kanno, Keita and Kohda, Masaya and Imai, Ryosuke and Koh, Sho and Mitarai, Kosuke and Mizukami, Wataru and Nakagawa, Yuya O.},
  journal = {Phys. Rev. Res.},
  volume = {8},
  issue = {2},
  pages = {023268},
  numpages = {27},
  year = {2026},
  month = {Jun},
  publisher = {American Physical Society},
  doi = {10.1103/dmn4-snfx},
  url = {https://link.aps.org/doi/10.1103/dmn4-snfx}
}
\newpage
\section*{Appendix A: Details of DMRG calculations} \label{app:dmrg}

For comparison, we include the ground-state energies of the beryllium dimer (Be$_2$) and water (H$_2$O), computed with the density-matrix renormalization group (DMRG), as
implemented using the \texttt{block2} package (accessed through its
\texttt{pyblock2} Python interface \cite{pyblock2}). One- and two-electron integrals were taken
from a PySCF restricted Hartree--Fock (RHF) calculation, and all calculations
were all-electron (no frozen core) and spin-adapted using SU(2) symmetry. CCSD(T) energies from PySCF were computed alongside as cross-checks. Each molecule was treated in both the cc-pVDZ and cc-pVTZ bases. The resulting active spaces
were: Be$_2$ --- 28 orbitals (cc-pVDZ) and 60 orbitals (cc-pVTZ); H$_2$O ---
24 orbitals (cc-pVDZ) and 58 orbitals (cc-pVTZ); with the two molecules having 8 (Be$_2$)
and 10 (H$_2$O) electrons.

An identical 16-sweep schedule was used in all calculations, ramping the bond dimension from 250 to 1000 (the maximum bond dimension for every run) while decreasing the
perturbative noise. The details are as follows:
\begin{lstlisting}
bond_dims = [250]*4 + [500]*4 + [1000]*8  # 16 sweeps, ramping M
noises    = [1e-4]*4 + [1e-5]*4 + [1e-6]*4 + [0.0]*4
thrds     = [1e-10]*16  # Davidson threshold
tol       = 1e-8        # sweep energy convergence
\end{lstlisting}
The DMRG ground-state energies are reported in
Table~\ref{tab:dmrg}.

\begin{table}[htbp]
  \centering
  \caption{Reference DMRG ground-state energies ($M = 1000$, SU(2)
  spin-adapted), with RHF and CCSD(T) references. All energies are in hartree, reported to 4 decimal places. All distances and coordinate values are reported to 3 decimal places.}
 \label{tab:dmrg}
  \centering
  \small
  \begin{tabular}{@{}ll>{\raggedright\arraybackslash}p{4.4cm}rrr@{}}
    \toprule
    Molecule & Basis & $R$ (\AA) & RHF & CCSD(T) & DMRG ($M{=}1000$) \\
    \midrule
    Be$_2$  & cc-pVDZ & 2.564 & $-29.1346$ & $-29.2345$ & $-29.2352$ \\
    Be$_2$  & cc-pVTZ & 2.385 & $-29.1320$ & $-29.2540$ & $-29.2546$ \\
    \midrule
    Molecule & Basis & Coordinates (\AA) & RHF & CCSD(T) & DMRG ($M{=}1000$) \\
    \midrule
    H$_2$O  & cc-pVDZ & O: $(0,\,0,\,0)$; \newline H: $(\pm 0.750,\,0.608,\,0)$
            & $-76.0262$ & $-76.2434$ & $-76.2440$ \\
    H$_2$O  & cc-pVTZ & O:$ (0,\,0,\,0)$; \newline H: $(\pm 0.754,\,0.588,\,0)$
            & $-76.0572$ & $-76.3458$ & $-76.3458$ \\
    \bottomrule
  \end{tabular}
\end{table}


\paragraph{Convergence note.} All DMRG calculations were run on an Apple M4 pro, multi-threaded across 8 threads. At the reported precision of four decimal places, all DMRG energies are fully converged with respect to bond dimension: the truncation errors lie three or more orders of magnitude below the last reported digit (the discarded weights $\lesssim 10^{-7}$, which correspond to errors of at most tens of $\mu$Ha). For the specific case of Be$_2$ in the cc-pVDZ basis set, we empirically verified this as well, by repeating the calculation at $M=2000$ and $M=3000$. Orbital reordering using the \texttt{fiedler} and \texttt{gaopt} algorithms also did not change the reported energy in this case.
\newpage
\section*{Appendix B: Convergence of H2O calculations} \label{app:H2O}
The convergence of the gradient during the VGSQD geometry optimization of H$_2$O presented in Table~\ref{tab:results} is shown in Figure~\ref{fig:h2oopt} below.
\begin{figure}[htbp]
\centering
\includegraphics[width=\linewidth]{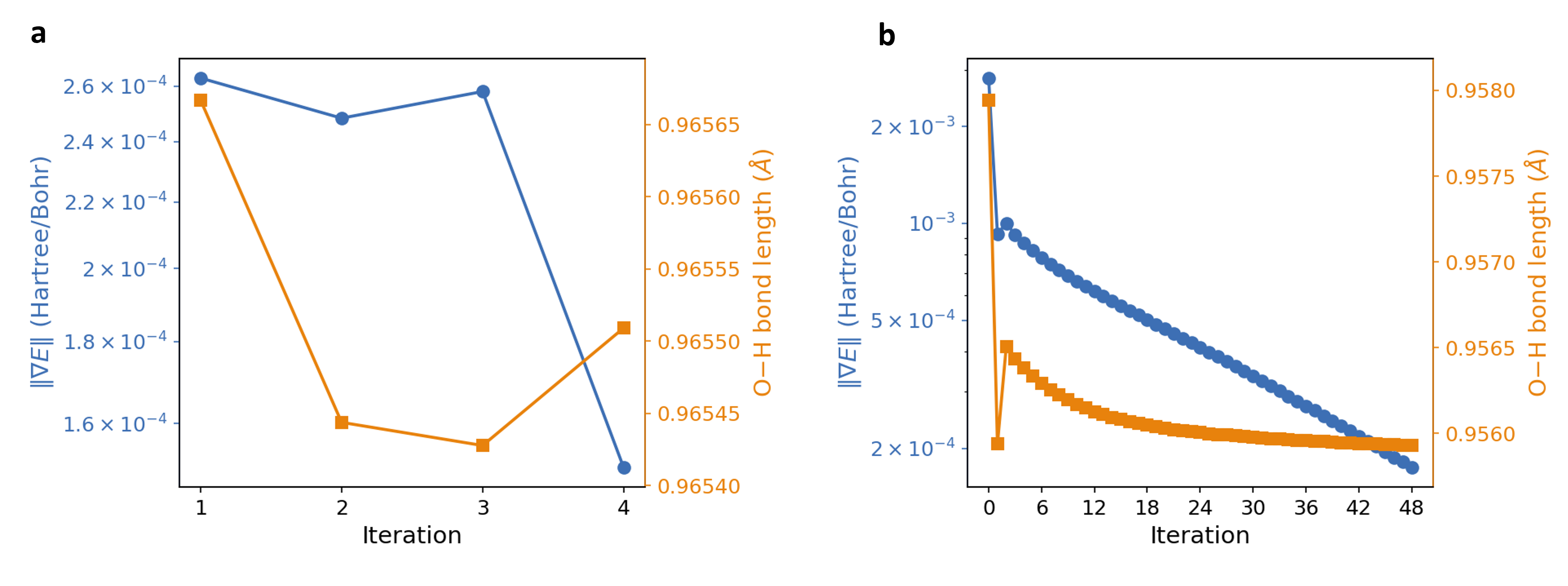}
\caption{Gradient convergence for the VGSQD geometry optimization of H$_2$O with (a) the cc-pVDZ basis set and (b) cc-pVTZ basis set.}
\label{fig:h2oopt}
\end{figure}
\newpage
\section*{Appendix C: Convergence of CI-NEB calculations} \label{app:NEBconv}
The convergence of the gradient and barrier height during the VGSQD CI-NEB calculation of the $\mathrm{H_2} + \mathrm{H} \rightarrow \mathrm{H} + \mathrm{H_2}$ reaction presented in Figure~\ref{fig:neb} is shown in Figure~\ref{fig:nebconv} below. We note that the molecular geometries of the initial path at which the VGSQD CI-NEB calculation is started were determined using a Hartree-Fock CI-NEB calculation.
\begin{figure}[htbp]
\centering
\includegraphics[width=\linewidth]{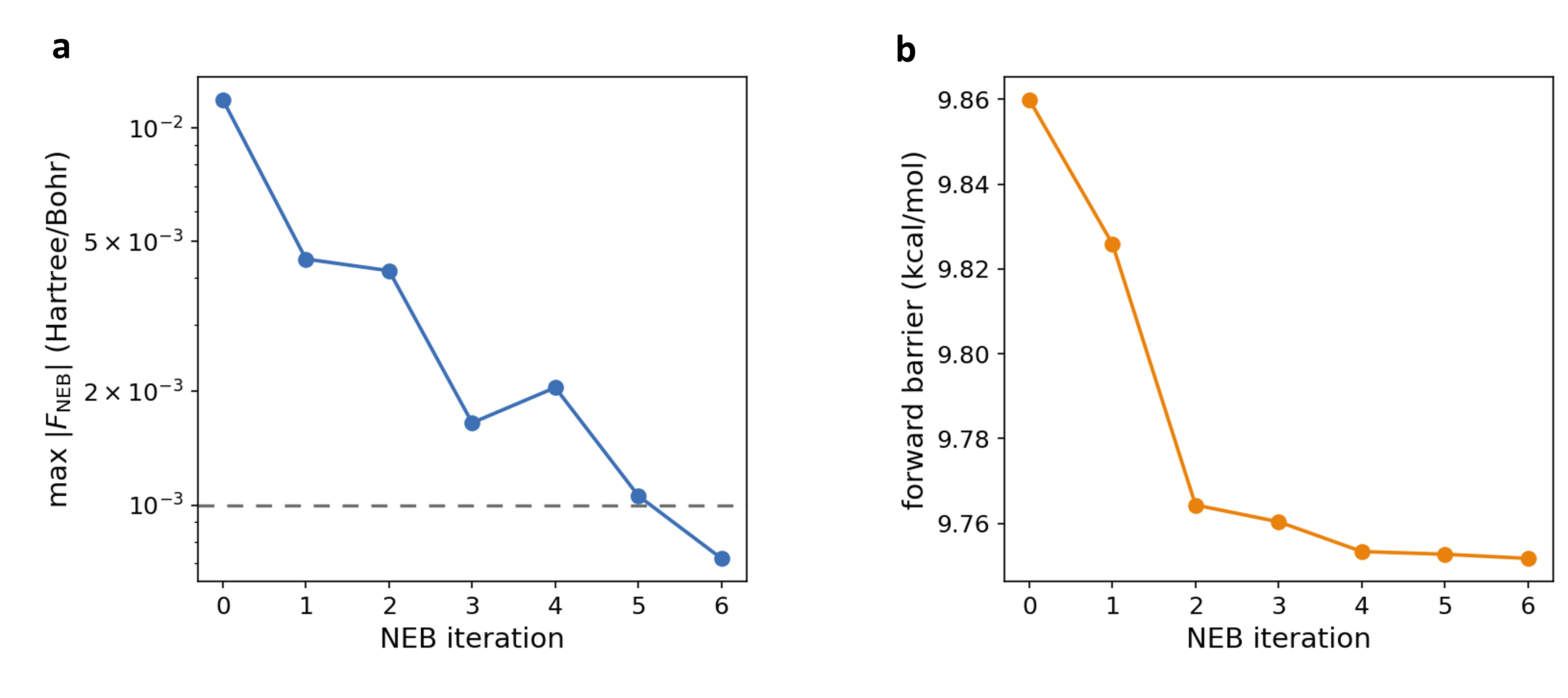}
\caption{Convergence of the CI-NEB calculation for $\mathrm{H_2} + \mathrm{H} \rightarrow \mathrm{H} + \mathrm{H_2}$ showing the convergence of the forces in (a) and of the barrier height in (b).}
\label{fig:nebconv}
\end{figure}

\end{document}